# Photochemically induced dynamic nuclear polarization of heteronuclear singlet order

Kirill F. Sheberstov[1,2], Liubov Chuchkova[2,3], Yinan Hu[1,2], Ivan V. Zhukov[4,5], Alexey S. Kiryutin[4,5], Artur V. Eshtukov[6], Dmitry A. Cheshkov[6], Danila A. Barskiy[1,2], John W. Blanchard[2], Dmitry Budker[1,2,7], Konstantin L. Ivanov[4,5], Alexandra V. Yurkovskaya[4,5]

1. *Institut für Physik, Johannes Gutenberg Universität-Mainz, 55128 Mainz, Germany*
2. *Helmholtz-Institut Mainz, GSI Helmholtzzentrum für Schwerionenforschung, 55128 Mainz, Germany*
3. *Faculté des Sciences Mirande, Université de Bourgogne, Dijon 21078, France*
4. *International Tomography Center SB RAS, Novosibirsk, 630090, Russia*
5. *Novosibirsk State University, Novosibirsk, 630090, Russia*
6. *State Scientific Research Institute of Chemistry and Technology of Organoelement Compounds, 105118, Moscow, Russia*
7. *University of California at Berkeley, California 94720-7300, USA*

## Abstract

Photochemically induced dynamic nuclear polarization (photo-CIDNP) is a method to hyperpolarize nuclear spins using light. In most cases, CIDNP experiments are performed at high magnetic field and the sample is irradiated by light inside a nuclear magnetic resonance (NMR) spectrometer. Here we demonstrate photo-CIDNP hyperpolarization generated in the Earth's magnetic field and under zero- to ultralow-field (ZULF) conditions. Irradiating a sample for several seconds with inexpensive light-emitting diodes produces a strong hyperpolarization of $^1$H and $^{13}$C nuclear spins enhancing the NMR signals by several hundred times. The hyperpolarized spin states at the Earth's field and in ZULF are different. In the latter case the state corresponds to singlet order between scalar-coupled $^1$H-$^{13}$C nuclear spins. This state has longer lifetime than the state hyperpolarized at Earth's field. The method is simple and cost-efficient and should be applicable to many molecular systems known to exhibit photo-CIDNP, including amino acids and nucleotides.

**Table of Contents Graphic**

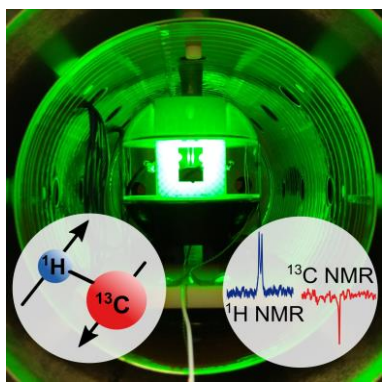

**Main text**

Nuclear magnetic resonance (NMR) is a widespread method to study composition of substances, molecular structure, molecular dynamics, and to perform noninvasive imaging. The main drawback of the method is its low sensitivity. Nuclear spins are only weakly polarized at equilibrium conditions. Even at strong magnetic fields of the order of 20 T, the level of polarization (defined as the relative population difference of spin states) at room temperature is of the order of $10^{-4}$-$10^{-5}$. This means that effectively only one molecule in 10,000 or even 100,000 contributes to the NMR signal, rendering the sensitivity notoriously low. One of the ways to boost NMR signal is to perform photochemically induced dynamic nuclear polarization (photo-CIDNP)[1], giving rise up to several 1000-fold signal enhancement[2]. Apart from hyperpolarizing the spins, CIDNP also enables studying short-lived radicals[3], obtaining information about their magnetic properties[4], and exploring mechanisms of chemical reactions[5]. Photo-CIDNP is also a suitable method to polarize biologically relevant molecules,[6] including amino acids[7], polypeptides and proteins[8], and nucleotides,[9] and to study their reactivity.

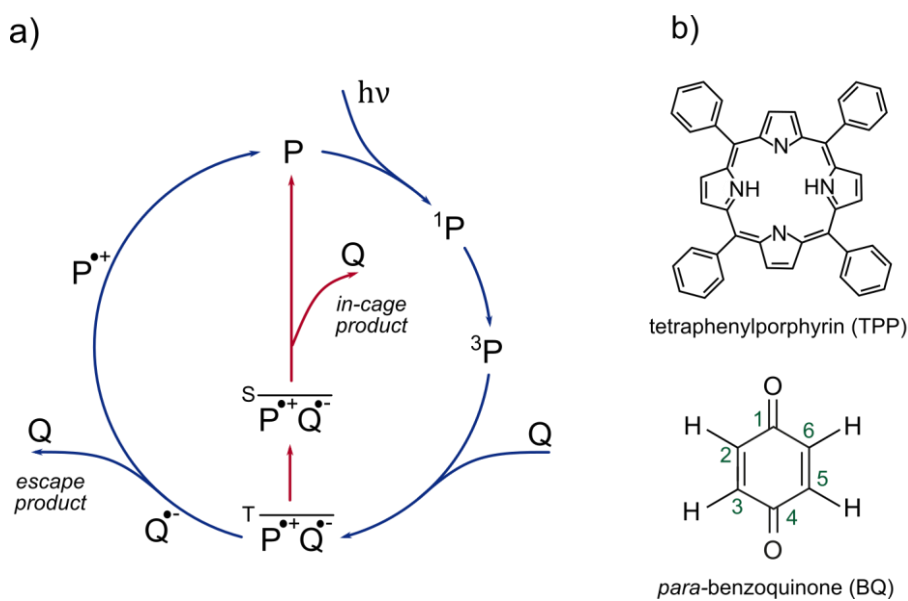

*Figure 1. (a) Radical-pair mechanism for a reversible reaction. The mechanism includes photoexcitation of the photosynthesizer P, followed by formation of a spin-correlated triplet radical pair, consisting of P· and Q· (Quencher) radicals, marked with an overbar. Triplet-singlet evolution in the radical pair is driven by hyperfine interactions and by the difference of the electronic Zeeman interaction with the magnetic field, thus providing nuclear spin selection leading to CIDNP. Triplet radical pairs do not react and form escape radicals and escape products; singlet radical pairs can undergo back electron transfer and form in-cage products enriched with nuclear spins that provide the fastest singlet-triplet conversion. (b) Molecular structures of compounds studied here. Tetraphenylporphyrin (TPP) was used as the photosynthesizer (electron donor) and para-benzoquinone (BQ) as the quencher (electron acceptor).*

Hyperpolarization in CIDNP is explained by a radical-pair mechanism[10] leading to spin selection in chemical reactions[11]. It is illustrated in Figure 1 for a reversible photoactivated chemical reaction with hyperpolarization produced in geminal radical pairs, which are formed in a single reaction event. The reaction stages include: (i) light excitation of the photosynthesizer molecule, *P*; (ii) formation of a spin-correlated radical pair between the exited *P* and a quencher, *Q*; (iii) singlet-triplet interconversion; (iv) spin-selective reactions of radical pairs. Zeeman and hyperfine interactions of the electrons cause conversion of the initial triplet state into the singlet state of the electron spins. This evolution depends on the nuclear spin states. (iv) Recombination or separation of the radical pair. In certain nuclear spin states interconversion is faster, and therefore the in-cage products (i.e., products of recombination of the primary radical pair) are enriched in these nuclear spin states. For other nuclear spin states the interconversion is slower, hence, these states are enriched in the escape products. In reversible processes, both in-cage and escape products recombine to the initial molecules *P* and *Q*. However, lifetime of the escape radicals is much longer than lifetime of the radicals that recombine in-cage. During this time the nuclear spins in escape radicals quickly relax due to efficient paramagnetic relaxation, while nuclear spin states in the in-cage products are preserved. This relaxation contrast leads to formation of the hyperpolarized in-cage products in geminal products. Nuclear polarization may also be strongly affected by the secondary radical pairs formed out of escape radicals (free pairs)[12].

Spin-selective recombination of radical pairs depends on the rate of the electron singlet-triplet interconversion, which, in turn, depends on the external magnetic field strength[13] and on the nuclear spin state. Whereas at high fields CIDNP corresponds to net polarization of spins parallel or anti-parallel to the magnetic field, at zero magnetic field net polarization goes to zero, but nuclear spin states are sorted with respect to their total spin[14]. Theory quantitively takes into account diffusion of the radicals, nuclear relaxation, and the evolution happening during field cycling[15], allowing one to carefully calculate magnetic field dependence of CIDNP[16].

The notion of "zero" magnetic field is relative, meaning that the strength of the residual field which may be considered as "zero" is different for different experiments. Zero field in CIDNP is determined by the spin dynamics in the radical pair (defining the spin flipping mechanism) and by the NMR parameters of the reaction product (defining the symmetry of the nuclear spin states). Usually, zero-field conditions for CIDNP are defined considering the spin dynamics in radical pairs and assuming that the electron Zeeman interaction with the external field is much weaker than hyperfine interactions with magnetic nuclei. Assuming that hyperfine couplings are of the order of 10 MHz and typical Zeeman interaction of an electron is around 30 GHz/T, we find that the relevant range corresponds to fields lower than 30 µT (nearly the strength of the Earth's magnetic field).

In turn, the zero- to ultralow-field (ZULF) regime for the reaction products corresponds to the case where the difference in Larmor frequencies of interacting nuclear spins is much smaller than the *J*-coupling between them. In the case of heteronuclear spins, this condition is met at magnetic fields of the order of 1 µT or lower[17]. In this work we report a study of photo-CIDNP in a $^1$H-$^{13}$C spin system generated at low and ultralow fields. We demonstrate that the polarization character significantly changes when the magnetic field is lowered from the Earth's field to approximately 1 µT.

ZULF-NMR methodology has experienced rapid progress within the last decade owing to development of sensitive atomic magnetometers[17]. Interest in ZULF-NMR techniques has arisen for several reasons. First, ZULF-NMR spectra are characterized by excellent spectral resolution as the NMR lines are not subject to any significant inhomogeneous broadening. Second, ZULF conditions are favorable[18] for polarization transfer among heteronuclei, e.g., for polarization transfer from protons to $^{13}$C or $^{15}$N nuclei. This property is frequently exploited in spin hyperpolarization experiments, notably, in experiments with parahydrogen[19]. Third, in some cases non-thermal spin order in heteronuclear systems is long-lived under ZULF conditions. An example of such a long-lived state is singlet order of two spin-1/2 heteronuclei, defined as the population imbalance between the singlet and triplet states of the pair. Such long-lived states are well documented for homonuclei[20] and recently long-lived order was also reported for $^1$H-$^{13}$C spin pairs at ultralow fields[21–23]. Here we show that photo-CIDNP can also generate long-lived heteronuclear hyperpolarization under ZULF conditions.

At ZULF, CIDNP can create population imbalance between states with different total nuclear spin, where all states belonging to the manifold with the same total spin have identical populations[14]. We refer to this polarization pattern as the zero-field multiplet effect. The simplest example of such a state is given by the singlet-triplet population imbalance for two coupled spins (with the three triplet states being equally populated). Previously, hyperpolarization of the singlet order by photo-CIDNP in low magnetic fields was observed for β-CH$_2$ protons in N-acetyl histidine[24] and tyrosine[25]. This spin order was long-lived, with a lifetime 45 times longer than the corresponding longitudinal relaxation time[26].

In the present work, we study photo-CIDNP of tetraphenylporphyrin (TPP, *P*) reacting with *para*-benzoquinone (BQ, *Q*). The molecular structures are shown in Figure 1b. This is a model system with strong CIDNP enhancement reported at high fields[27–31], allowing us to study CIDNP of compounds with natural abundance of $^{13}$C. CIDNP can be initiated with green light, allowing usage of inexpensive and powerful LEDs[32]. The photoreaction between TPP and BQ is reminiscent to some stages of photosynthesis: there, an electron is transferred from a photoactivated porphyrin fragment, which is part of chlorophyll, to a quinone. In our case, according to previous studies, radical pairs are formed by TPP and BQ radicals, both of which are protonated[30] – a conclusion supported by the fact that CIDNP is observed only in the presence of an acid.

The experimental protocol used here is illustrated in Figure 2. The sample was placed in a specially designed chamber and irradiated with green light produced by LEDs positioned around the NMR tube. Irradiation occurred inside a four-layer magnetic shield. A Helmholtz coil was installed around the sample to generate magnetic field, allowing us to study the field dependence of CIDNP. The light was switched on for 5 s to initiate photo-CIDNP (Figure 2, stage i). The composition of the sample, timing, and light intensity were optimized to result in stable and strong CIDNP reproducible for at least 100 irradiations per sample. To study the lifetime of the hyperpolarized state, the sample was held in the irradiation chamber for a variable delay after the light illumination (Figure 2, stage ii). Subsequently, the sample was transferred with a robotic arm into a 60 MHz benchtop NMR spectrometer for detection (Figure 2, stages iii and iv). For each spectrum shown, this procedure was repeated several times to improve signal-to-noise ratio by averaging. A more detailed description of the experiment is given in the Supporting Information.

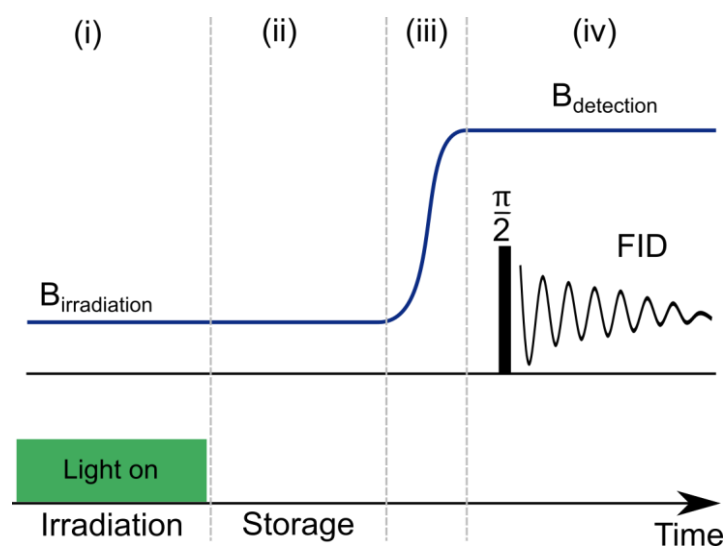

*Figure 2. (a) Experimental protocol to study CIDNP at low field. (i) Sample at low magnetic field; light on for 5 s. (ii) Optional variable-time storage to study relaxation. (iii) Transfer to NMR spectrometer. (iv) RF pulse to generate free induction decay (FID). After Fourier transform of the FID NMR spectrum is obtained.*

The $^1$H NMR spectrum of the sample at thermal equilibrium at 1.4 T contains signals of TPP (0.5 mM) and BQ (5 mM) and the signals of residual protons in the solvents (deuterated chloroform and deuterated acetic acid were used with volume ratio of 70% to 30%). The $^1$H spectrum is shown at the top row I in Figure 3a, and all signals come from $^{12}$C-isotopomers. The sensitivity was not high enough to detect any lines coming from $^{13}$C-isotopomers at natural abundance in thermal equilibrium. There were no visible signals in the $^{13}$C channel (Figure 3b, I). This changes in the case of low-field CIDNP hyperpolarization – additional lines appear in $^1$H and $^{13}$C NMR spectra (Figure 3a, II and III and Figure 3b, III), corresponding to $^{13}$C-isotopomers of BQ. To verify this, we studied highly concentrated samples of BQ (0.4 M) and assigned the signals. Total lineshape analysis[33] was performed for the $^1$H and $^{13}$C spectra and all *J*-couplings in 2-$^{13}$C-BQ and 1-$^{13}$C-BQ were determined (see Supporting Information). In 2-$^{13}$C-BQ the value of the direct $^1J_{HC}$-coupling is approximately 170 Hz, the second largest *J*-coupling is between the two neighboring protons and is approximately 10 Hz. In the 1-$^{13}$C-BQ isotopomer the largest $^3J_{HC}$-coupling slightly exceeds 10 Hz.

The lines corresponding to the 2-$^{13}$C-BQ isotopomer appear in the $^1$H NMR spectrum as two satellites of the main $^{12}$C-BQ line. These lines are strongly enhanced by photo-CIDNP. Estimation using the integral intensities relative to the $^{12}$C-BQ line shows that the enhancement factor exceeds 200 times. When CIDNP is performed at Earth's field (Figure 3a, II), the $^{13}$C satellites exhibit an antiphase pattern, with the left component appearing as absorptive and right component as emissive signal. When the sample is irradiated at ZULF, the appearance of the $^1$H spectrum changes, and only the left component is pronounced (Figure 3a, III).

An advantage of $^{13}$C NMR is the inherently higher resolution due to larger range of possible chemical shifts. Since photo-CIDNP in ZULF produces net magnetization of $^{13}$C spins (as discussed below), it is possible to perform proton decoupling without cancelling the signal and thus significantly improving the signal-to-noise ratio (Figure 3b, III). Signals produced after photo-CIDNP in the Earth field on the other hand were canceled out (Figure 3b, II). It is possible to acquire them without proton decoupling, however, we could not see any signals in 50 acquisitions. The $^{13}$C NMR spectrum revealed that not only 2-$^{13}$C-BQ but also the 1-$^{13}$C-BQ isotopomer is hyperpolarized by photo-CIDNP. The line corresponding to 1-$^{13}$C-BQ in the $^1$H spectrum normally overlaps with the signal of $^{12}$C-BQ, but for the ZULF case this central line was observed only if additional measures were taken to provide adiabaticity of the transfer through the magnetic shield (see Supporting Information). The integral intensity of the $^{13}$C line corresponding to 2-$^{13}$C-BQ is two times stronger than the line of 1-$^{13}$C-BQ, signifying that the efficiency of photo-CIDNP is identical for them (there are twice as many of the former isotopomers as the latter).

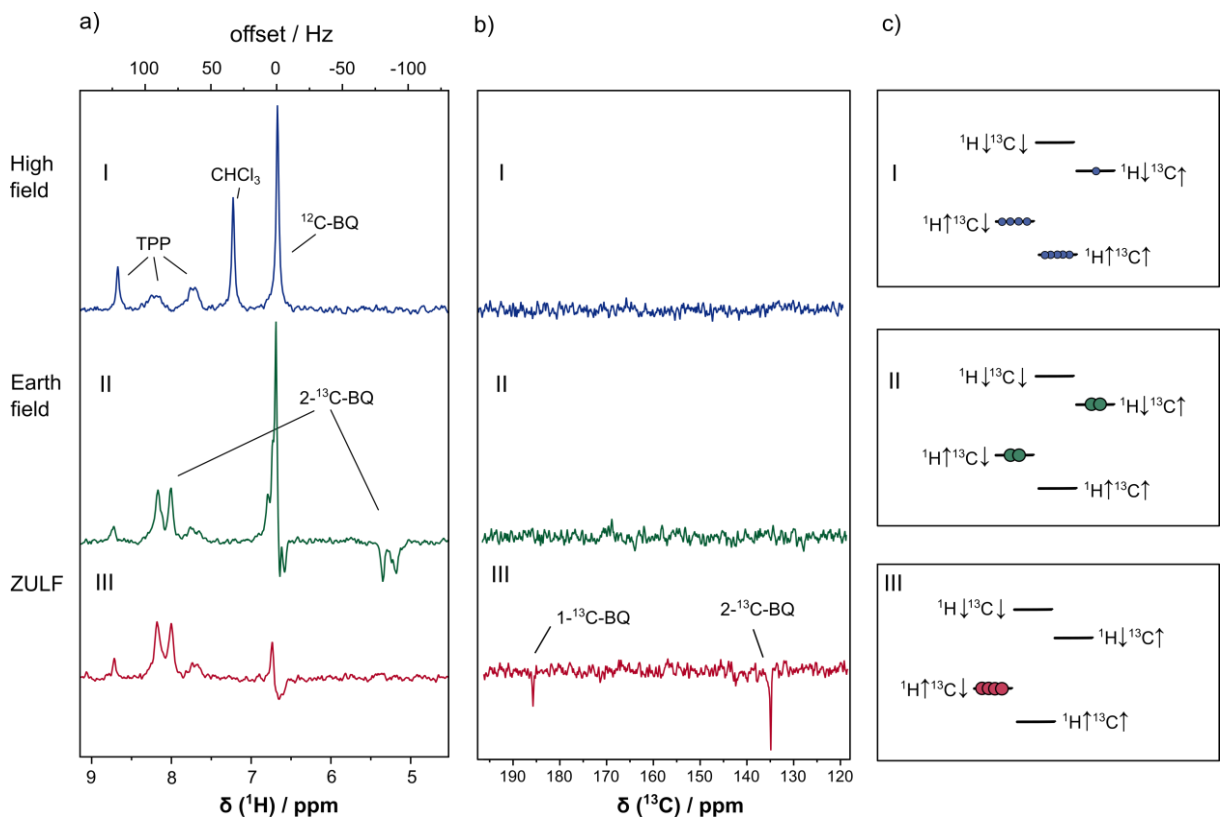

*Figure 3. (a) 60 MHz $^1$H NMR spectra obtained after averaging of 10 acquisitions. (b) 15 MHz $^{13}$C NMR spectra obtained by averaging of 50 acquisitions. $^{13}$C spectra were acquired with broadband $^1$H decoupling to increase the signal-to-noise ratio. The phase of the $^{13}$C spectra was determined using a reference $^{13}$C-enriched methanol spectrum with identical acquisition parameters. (c) High-field spin-state populations in (a) and (b). The spin-up or spin-down states of $^1$H and $^{13}$C spins are shown with arrows; excess population of states is shown with filled circles. The three rows I, II, and III correspond to differently prepared spin states: (I) thermal equilibrium at high field, (II) photo-CIDNP generated at Earth field (40 uT), and (III) photo-CIDNP generated at ZULF (1 uT).*

All presented results can be explained by analyzing the nuclear spin states formed during the photo-CIDNP process. A detailed theoretical treatment is beyond the scope of the present work; here we use the results of earlier considerations[14,16]. The crucial assumption is that in the BQ radical, hyperfine couplings are nonzero for all the $^1$H and $^{13}$C nuclei. This agrees with electron paramagnetic resonance experiments that showed that the $^1$H hyperfine coupling for the BQ cation radical is around 6.3 MHz[34] and, for the BQ anion radical, it is around –6.7 MHz[35]. The hyperfine couplings with $^{13}$C in benzoquinones are larger by approximately an order of magnitude[36]. In this situation, CIDNP theory predicts that "multiplet" polarization should be generated at zero field, meaning that a population imbalance is created between the singlet and triplet states of the $^1$H-$^{13}$C spin pair (where the populations of the triplet states are equal). After recombination of the radical pair, the non-thermally polarized state is projected onto the eigenstates of the nuclear spin system of BQ. The eigenstates of the nuclear spin system strongly depend on the magnetic field.

In the following we simplify the spin system of 2-$^{13}$C-BQ and ignore the presence of the three protons that are not directly bonded to the $^{13}$C spin. The full spin system is considered in the Supporting Information and it is shown that the conclusions remain the same.

The nuclear spin eigenstates at Earth's field nearly coincide with the Zeeman product states. This is because the difference in the precession frequencies of the $^1$H and $^{13}$C spins at Earth field is much larger than the *J*-coupling between them (the high-field regime). After adiabatic transfer of the sample into the high field of the NMR spectrometer, the eigenstates remain almost unchanged. Projection of the hyperpolarized singlet order onto the Zeeman states results in population imbalance between the states with total z-projection of the nuclear spins of 0 and ±1. Therefore, spin-state populations are distributed as shown in the middle row of Figure 3c, namely, the |↑↓⟩ and |↓↑⟩ states are overpopulated.

When photo-CIDNP is performed under ZULF conditions, the eigenstates of the directly bound $^1$H-$^{13}$C spin pair are the singlet and triplet states. Therefore, the hyperpolarized singlet order remains unchanged when it is projected onto the eigenbasis of the spin system of the diamagnetic reaction product. After an adiabatic transfer of the sample, the populations follow the instantaneous time-dependent eigenstates; consequently, at the detection field the spin state populations are distributed as depicted at the bottom row of Figure 3c: only one of the four states is populated, namely, the |↑↓⟩ state, which is correlated with the singlet state at zero field.

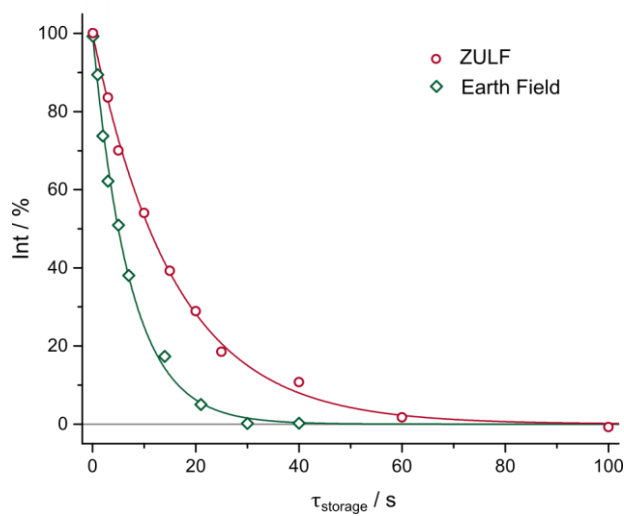

*Figure 4. Decay of the $^1$H signals of 2-$^{13}$C-BQ molecules hyperpolarized at ZULF and Earth's field. Each experimental point corresponds to averaged integral, as in Figure 3a, II and III. The solid lines represent fitting to an exponential decay. The extracted relaxation times are 15.8 ± 0.5 s for the ZULF case and 7.2 ± 0.3 s for the Earth-field case.*

The heteronuclear singlet order prepared at ZULF conditions can be long-lived[21–23]. To verify this, we performed additional experiments to determine the lifetime of the $^1$H-

$^{13}$C singlet order in the 2-$^{13}$C-BQ molecule. For the degassed sample, the singlet order lifetime reaches ~16 s, which is about two times longer than the lifetime of the spin state hyperpolarized at Earth's field (Figure 4).

There are many other molecules containing directly bonded $^{13}$C and $^{1}$H spins, which can be potentially hyperpolarized by CIDNP in ZULF-NMR experiments. The general trend is that the stronger CIDNP is on protons, the stronger it is on carbon spins[37]. Directly bonded $^{13}$C-$^{1}$H spin pairs are usually strongly coupled in ZULF as the corresponding *J*-couplings are at least an order magnitude greater than all other $^{1}$H-$^{1}$H or $^{1}$H-$^{13}$C *J*-couplings[38]. This renders the eigenstates of the $^{1}$H-$^{13}$C pair in ZULF to be close to the true singlet-triplet states. We expect that ZULF photo-CIDNP is, indeed, a general way to hyperpolarize heteronuclear long-lived singlet states of directly bonded $^{1}$H-$^{13}$C spin pairs. These considerations make photo-CIDNP a promising hyperpolarization method in ZULF-NMR experiments.

**Acknowledgments**

K.S. and L.C. are grateful to Oleg Tretiak for useful discussions. K.S. acknowledges support by Internal University Research Funding Stufe I at the JGU Mainz. D.B. acknowledges support by the Cluster of Excellence Precision Physics, Fundamental Interactions, and Structure of Matter (PRISMA+ EXC 2118/1) funded by the DFG within the German Excellence Strategy (Project ID 39083149). Y.H. acknowledges German Federal Ministry of Education and Research (BMBF) within the Quantumtechnologien program (FKZ 13N14439 and FKZ 13N15064). D.A.B. thanks Alexander von Humboldt Foundation in the framework of the Sofja Kovalevskaja Award. The Novosibirsk team acknowledges support from the Russian Science Foundation (grant No. 20-63-46034). This project has received funding from the European Union's Horizon 2020 research and innovation programme under the Marie Skłodowska-Curie grant agreement No 766402.

**Supporting Information**

1. Setup description

The setup for photo-CIDNP is shown in Figure S5. The sample was irradiated inside a 4-layers magnetic shield (Twinleaf, MS-2). Irradiation chamber was designed to accommodate NMR sample surrounded by up to 6 LEDs and a Hemholtz coil. It was printed out of acrylonitrile-butadiene-styrene using Ultimaker$^2$ Extended+ 3D printer. 4 LEDs (LED engine, LZ4-00G108-0000) with standard star MCPCB were used for irradiation. Only 2 LEDs are shown in Figure S5, 2 more were installed in perpendicular direction to irradiate the sample from all sides. To avoid overheating, each of the LED was glued using epoxy glue (Loctite) to an aluminum heatsink (Ohmite, SV-LED-314E). Additionally, the temperature was stabilized using flow of dry air blown from below the sample allowing to keep its temperature below 30 °C.

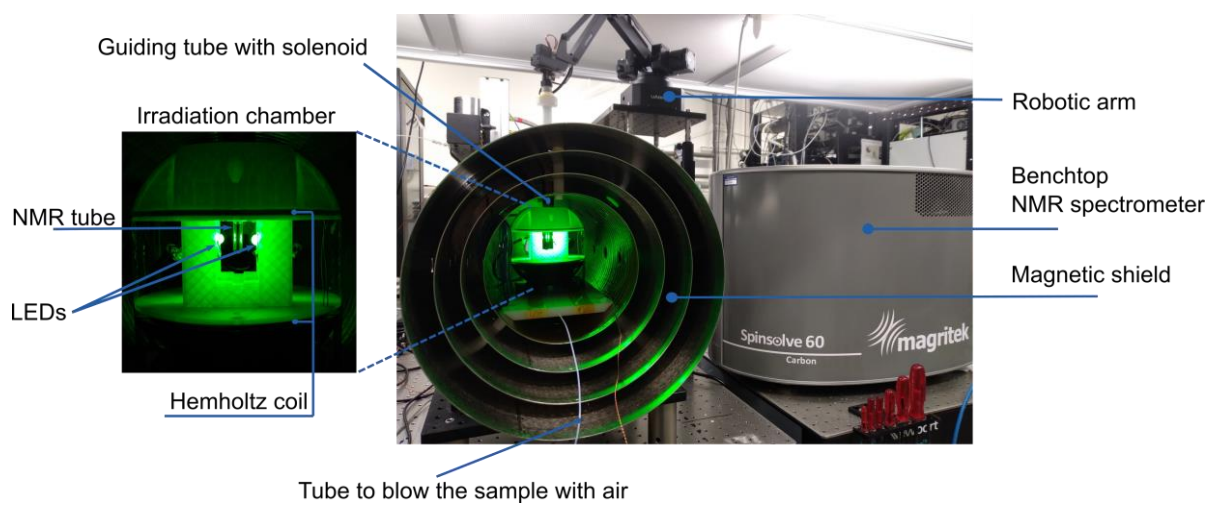

*Figure S5. Photo of the developed setup for the low-field photo-CIDNP.*

2 sets of 2 LEDs (connected in series) were connected to the two outputs of a power supply (RS PRO bench power supply, 123-6467) through programmable relays (Mitsubishi FX3S). Light intensity was controlled by current, with the optimal current value found to be around 100 mA. This current corresponded to 0.2 W of the output light intensity per LED measured at 515 nm; approximately 0.1 W of light power from each LED reached the sample (distance of 5 mm). Light intensity was measured using an optical power meter from Thorlabs (PM100D).

To minimize magnetic noise induced by the current in LEDs, they were connected to the power source using a twisted pair cable. The induced magnetic field was measured using a fluxgate magnetometer (Bartington Instruments, Mag-01H). LEDs themselves are slightly magnetic and even without any current induce a residual field, which was below 500 nT. Turning on the current increased the field; the measured field was less than 1 uT.

To introduce an adiabatic change of the magnetic field, the sample was transferred from the magnetic shield through a solenoid. The solenoid provided a magnetic field of 40 uT. Additionally, in ZULF experiments, before sample transfer the magnetic field around the sample was gradually increased (using a Helmholtz coil); subsequently, the field inside solenoid was turned on. The field strength was linearly increased from 0 to 40 uT in 1 s. To generate such a field ramp, we used an analogue output of a Red Pitaya microcontroller, which was further amplified (by AE tecron 7224) into linearly swept current through the Helmholtz coil. These additional measures are not necessary to observe the NMR lines corresponding to 2-$^{13}$C-BQ. However, the NMR lines corresponding to 1-$^{13}$C-BQ were not always possible to observe without these measures.

In order to run hundreds of experiments and to transport the sample in a reproducible way we used a robotic arm (Ufactory, Uarm Swift Pro). The arm took NMR samples using a "suction head": it picked up the sample from the NMR spectrometer and dropped it there. The motion from the ZULF chamber to the NMR spectrometer required 4.23 ± 0.03 s. After the sample was placed inside the spectrometer a trigger signal was sent to start the acquisition. NMR spectra were acquired using a Magritek Spinsolve 60 MHz spectrometer equipped with a $^{13}$C channel. The entire sequence of events was programmed using a Python code, allowing synchronization of motion of the arm, switching the LED state, varying the magnetic field in the coil and solenoid and NMR acquisition.

The intensity of photo-CIDNP strongly depends on the composition of the sample. Preliminary adjustments were done by acquiring high-field photo-CIDNP data using $^1$H line of $^{12}$C-BQ (not shown here). It was established that a mixture of CDCl$_3$ (70%) and CH$_3$OH (30%) allowed the highest enhancements. Concentration of TPP is mostly determined by absorption of light; hence, for optimization of the optical density we used a spectrophotometer (PerkinElmer, Lambda +1050). We have found that in this case the TPP concentration of 0.5 mM had provided the highest photo-CIDNP intensity, corresponding to the absorbance of 90% of the light in 2 mm of the solution. Finally, the BQ concentration was varied. The concentration ratio of 1 to 10 (TPP to BQ) appeared to be optimal in the ZULF case. Thus, after optimization the samples contained 0.5 mM TPP, 5 mM BQ diluted in a mixture of CDCl$_3$ (70%) and CH$_3$OH (30%). All chemicals were purchased from Sigma-Aldrich. All samples were preliminary degassed by bubbling the N$_2$ gas through a 5mm NMR tube for 5 minutes.

2. Stability of CIDNP

Photo-CIDNP of TPP with BQ is not a fully reversible process; consequently, slow degradation of the sample occurs. We performed control experiments to estimate the rate of the degradation. A typical depletion curve is shown in Figure S6. For a series of 100 ZULF photo-CIDNP experiments (0.5 mM TPP, 0.1W/LED, 5 s of irradiation) we observed the following trend: 10 first experiments exhibit slightly higher intensity of photo-CIDNP and then no significant change of the signal intensity is observed for at least 90 more experiments.

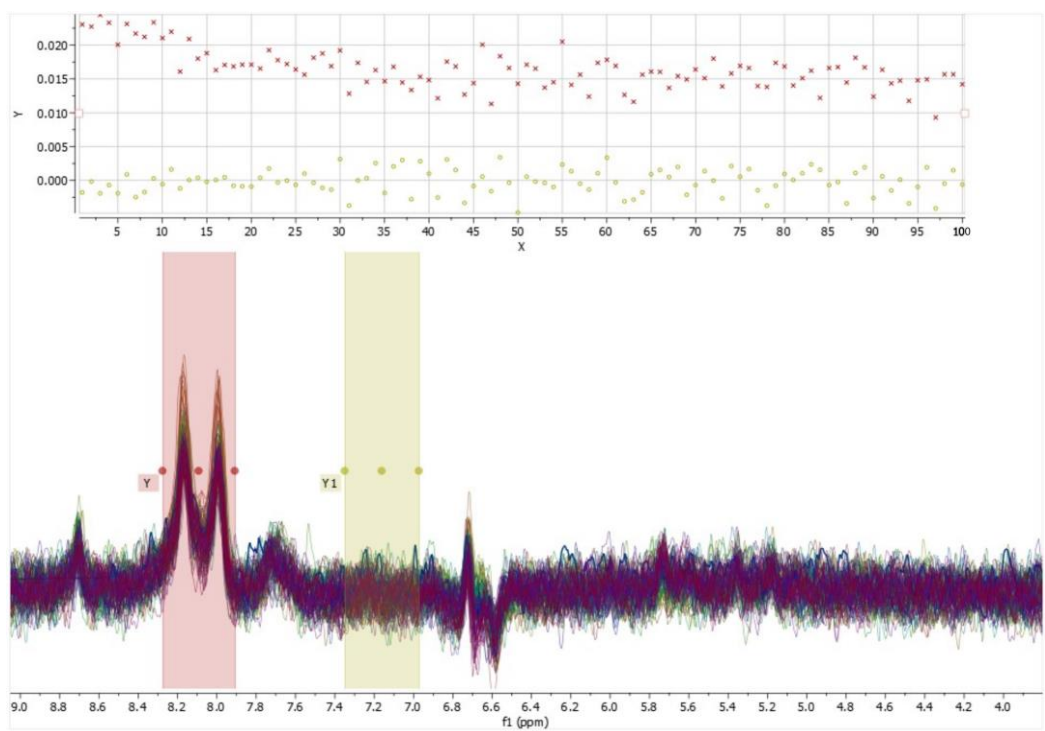

*Figure S6. Stability of photo-CIDNP in ZULF experiment. 100 overlapped $^1$H NMR spectra are shown at the bottom. Each spectrum obtained after single shot experiment. The signals of 2-$^{13}$C-BQ were integrated (red region) as well as a noise region for comparison (yellow region). The integral values as function of the experiment number are shown at the top.*

### 3. Lifetime measurements

Determination of photo-CIDNP stability was particularly important for measuring the lifetime of the hyperpolarized states generated in ZULF and in the Earth field (see Figure 4 of the main text). After confirming that photo-CIDNP is stable, we measured the lifetime kinetics (see Figure 2 of the main text) using a single sample. We introduced a sequence of relaxation delays (each set contained 10 points). For the ZULF case these timings were: (0.1, 3, 5, 10, 15, 20, 25, 40, 60, 100) s; for the Earth field case they were (0.1, 1, 2, 3, 5, 7, 14, 21, 30, 40) s. The entire range was recorded from the beginning to the end and this was repeated for 10 times to improve the signal-to-noise-ratio. Therefore, the lifetime of the photo-generated state was measured using the same sample in 100 experiments. We have also found a good reproducibility of photo-CIDNP for different samples. The final relaxation curve was averaged between several samples.

### 4. Analysis of high-resolution NMR spectra of benzoquinone

High-resolution $^1$H and $^{13}$C NMR of BQ in $C_6D_6$ (10 mM) spectra were acquired at 16.4 T ($^1$H Larmor frequency of 700 MHz). Total lineshape analysis was performed to find the *J*-couplings in $^{13}$C-BQ molecules using ANATOLIA software[1]. Both $^{13}$C and $^1$H spectra were fitted simultaneously to determine the best fitting parameters. As previously, the analysis was performed in several steps by gradually removing Lorentzian broadening of the experimental data and subsequently refining the spin system parameters[1]. Comparison of the experimental and calculated NMR spectra for 2-$^{13}$C-BQ is shown in Figure S7.

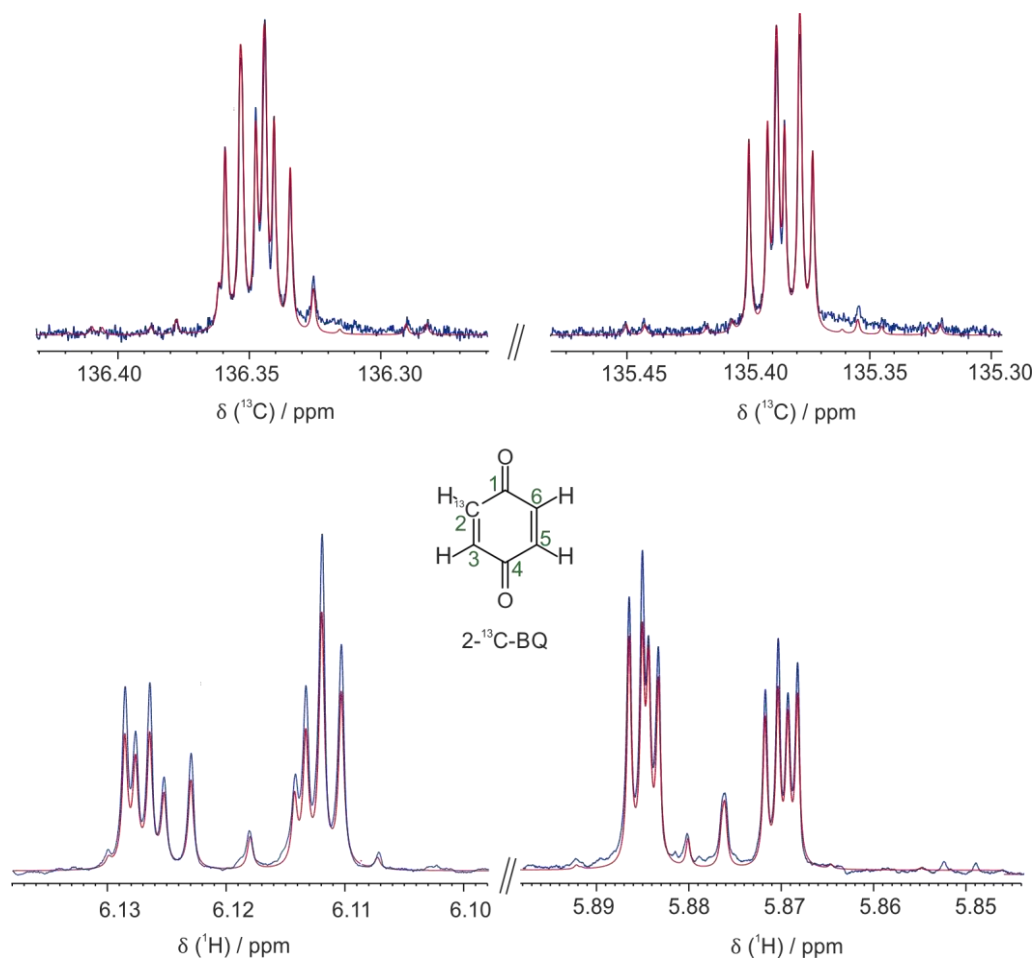

Figure S7. Total lineshape analysis of NMR signals corresponding to 2-$^{13}$C-BQ (16.4 T, 10 mM of BQ in $C_6D_6$). $^1$H and $^{13}$C NMR spectra were analyzed together. Results of the calculation (shown in red) are overlaid with the experimental spectra (shown in blue).

The results of this analysis for the 2-$^{13}$C-BQ isotopomer are presented in Table S1. Interestingly, all protons have a detectable isotopic shift induced by the presence of $^{13}$C nuclei. This caused a clear asymmetry of the NMR spectra; consequently, without variation of these parameters it was not possible to achieve a good agreement between the calculated and experimental spectra.

Table S1. Spin system parameters of 2-$^{13}$C-BQ determined from the simulation. The standard deviation was less than 0.03 Hz for each parameter.

| δ (ppm) | | ν (Hz) | | J (Hz) | |
|---|---|---|---|---|---|
| $\delta_{C2}$ | 135.868 | $\nu_{C2}$ | 23922.72 | $^4J_{H2-H6} = {}^4J_{H3-H5}$ | 2.59 |
| $\delta_{H2}$ | 5.998 | $\nu_{H2}$ | 4200.34 | $^3J_{H2-H3} = {}^3J_{H5-H6}$ | 10.16 |
| $\delta_{H3}$ | 5.999 | $\nu_{H3}$ | 4200.74 | $^5J_{H2-H5} = {}^5J_{H3-H6}$ | 0.00 |
| $\delta_{H5}$ | 6.001 | $\nu_{H5}$ | 4201.80 | $^1J_{H2-C2}$ | 168.70 |
| $\delta_{H6}$ | 6.001 | $\nu_{H6}$ | 4201.80 | $^2J_{H3-C2}$ | -0.85 |

| | | $^3J_{H6-C2}$ | 4.52 |
| | | $^4J_{H5-C2}$ | -0.73 |

$^1$H NMR spectra of 1-$^{13}$C-BQ isotopomer were not detected, because we used BQ with natural abundance of $^{13}$C spins. In this case a much stronger $^{12}$C-BQ line overlaps with the signals of 1-$^{13}$C-BQ. However, it is reasonable to assume that $J_{HH}$-couplings are the same in the 1-$^{13}$C-BQ and in 2-$^{13}$C-BQ isotopomers, ignoring possible minor isotopic effects. These values were given for the ANATOLIA for the input and fixed, thus making possible to analyze $^{13}$C spectrum of 1-$^{13}$C-BQ. This procedure provides a good agreement of the calculation and experiment (Figure S8).

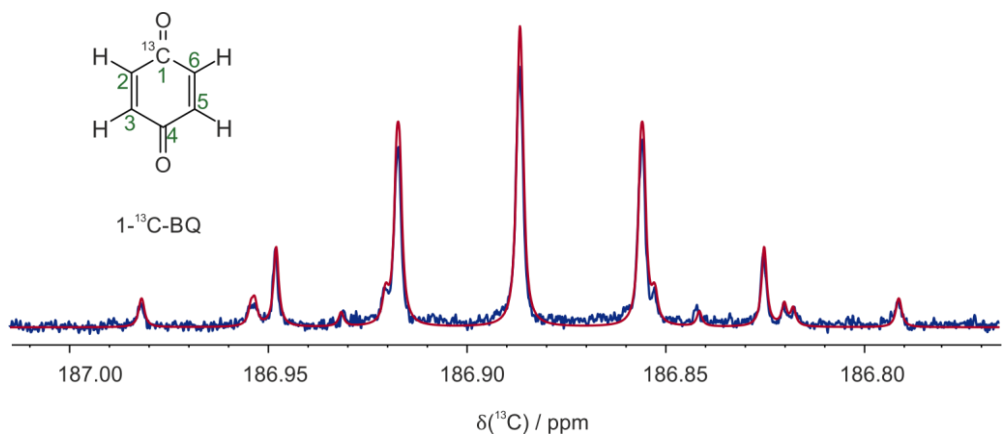

*Figure S8. Total lineshape analysis of the $^{13}$C NMR signals corresponding to 1-$^{13}$C-BQ (16.4 T, 10 mM of BQ in $C_6D_6$). J-couplings between the $^1$H spins were fixed to the values determined for the 2-$^{13}$C-BQ isotopomer. The result of the calculation (shown in red) is overlaid with the experimental spectrum (shown in blue).*

The parameters for the 1-$^{13}$C-BQ isotopomer extracted in this way are presented in Table S2.

*Table S2. Determined spin system parameters of the 1-$^{13}$C-BQ. Standard deviation was less than 0.03 Hz for each parameter.*

| δ (ppm) | | ν (Hz) | | J (Hz) | |
|---|---|---|---|---|---|
| $δ_{C1}$ | 186.885 | $ν_{C1}$ | 32905.37 | $^4J_{H2-H6} = {}^4J_{H3-H5}$ | 2.59 |
| $δ_{H2,H6}$ | 6.001 | $ν_{H2,H6}$ | 4201.87 | $^3J_{H2-H3} = {}^3J_{H5-H6}$ | 10.16 |
| $δ_{H3,H5}$ | 6.001 | $ν_{H3,H5}$ | 4201.87 | $^5J_{H2-H5} = {}^5J_{H3-H6}$ | 0.00 |
| | | | | $^3J_{H3-C1} = {}^3J_{H5-C1}$ | 10.47 |
| | | | | $^2J_{H2-C1} = {}^2J_{H6-C1}$ | 0.34 |

5. Photo-CIDNP in ZULF, consideration of the full spin system of 2-$^{13}$C-BQ

The extracted values of *J*-couplings were used to calculate the nuclear-spin energy levels of 2-$^{13}$C-BQ in zero magnetic field (Figure S9). The dominant interaction of the nuclear spins of 2-$^{13}$C-BQ in zero field is the *J*-coupling between the directly coupled $^{13}$C and $^1$H spins, which is ≈170 Hz being more than 10 times greater than all other interactions.

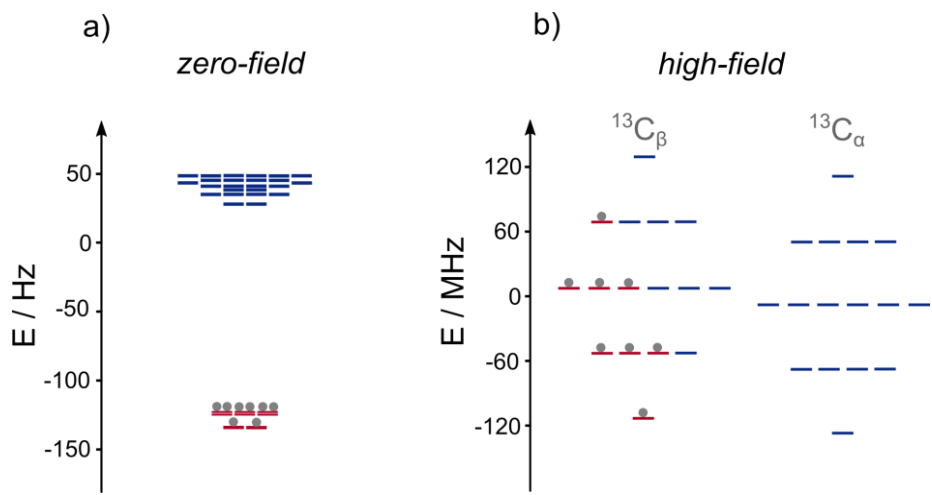

*Figure S9. Nuclear spin energy levels of 2-$^{13}$C-BQ in zero-field (a) and in a high field of 1.4 T (b). Photo-CIDNP in zero-field produces singlet order in the directly bonded $^{13}$C-$^1$H spin pair, which is illustrated as over-population of the lower energy levels in (a). Upon adiabatic transfer the overpopulated states can be brought to the high-field. The energy levels at high field are color-coded: red color is used for the levels that are correlated with the $^{13}$C-$^1$H singlet state at zero-field and the blue color is used for the levels that are correlated with the triplet states.*

We have simulated the experiments where detection is performed in the high field of an NMR spectrometer. Figure S9b shows the energy levels of $^{13}$C-BQ nuclear spins in a 1.4 T magnetic field. Here we consider an adiabatic transport of the hyperpolarized spin system from zero to the high-field. This procedure conserves the initial population of the levels upon variation of the magnetic field. The color of a level in Figure S9b reflects its correlation with the zero-field states. Consequently, the states, which are correlated with the singlet state, are overpopulated with respect to all the other states. Knowing the distribution of the spin populations before detection we calculated the $^1$H NMR and $^{13}$C spectra (Figure S10). There are three cases, corresponding to the initial state at thermal equilibrium (bottom row Figure S10), two-spin order hyperpolarized by photo-CIDNP in the Earth magnetic field (middle row Figure S10), and singlet order hyperpolarized by photo-CIDNP in ZULF (top row Figure S10). Appearance of the simulated spectra reproduces that of the experimental spectra shown in Figure 3 of the main article.

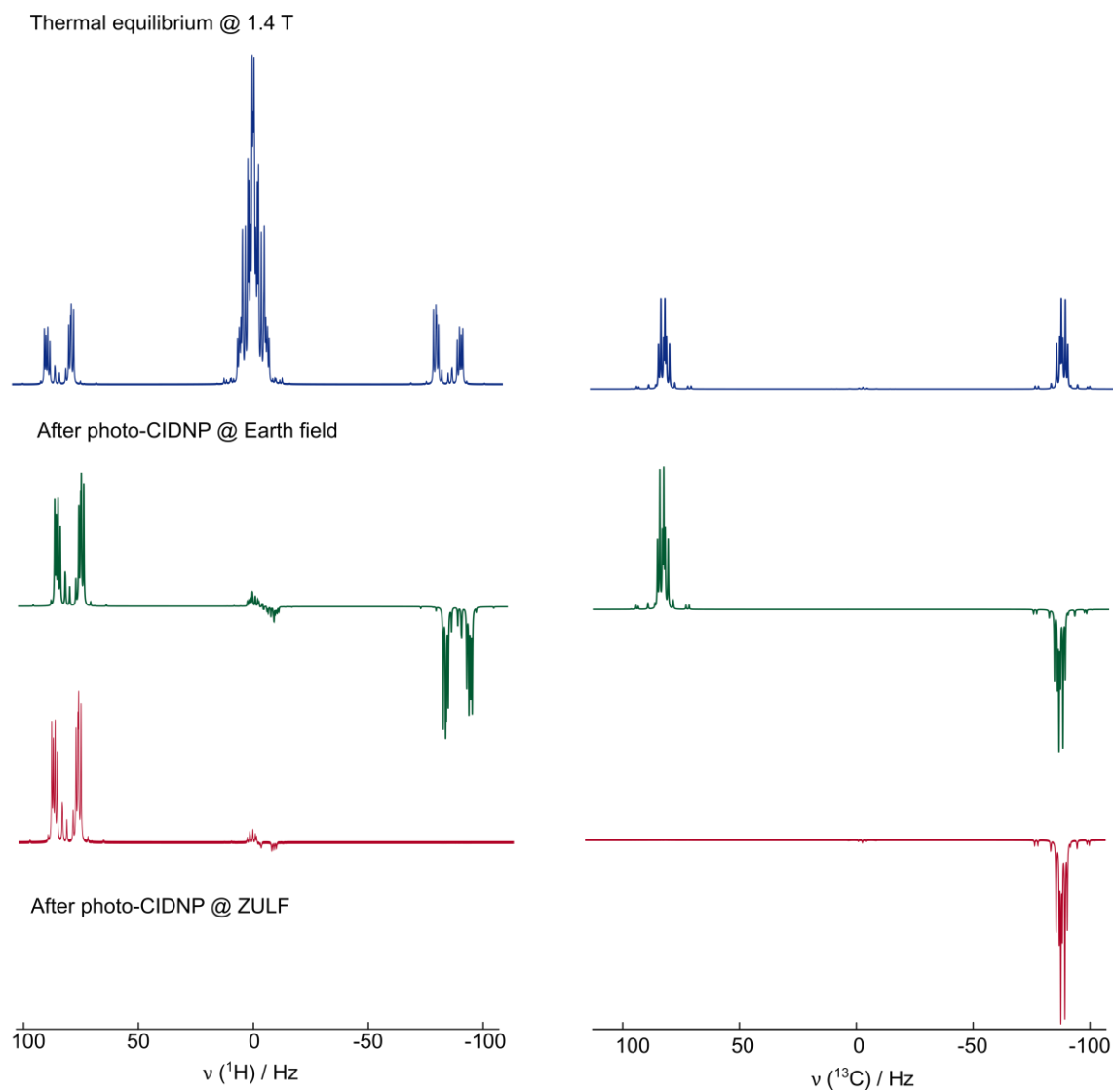

*Figure S10. Simulated $^1$H (left) and $^{13}$C (right) NMR spectra of 2-13C-BQ for different states before acquisition. In the top row, the state corresponds to thermal equilibrium at high field. In the middle row, the state corresponds to the state hyperpolarized by CIDNP at the Earth field (two-spin order). In the bottom row, the state corresponds to singlet order, produced by photo-CIDNP in ZULF.*